\documentclass[12pt]{article}
\usepackage{amssymb,graphicx}
\usepackage{epstopdf}
\usepackage{tabularx}
\usepackage{array}
\usepackage{subcaption}
\usepackage{amsmath,amsfonts}
\usepackage{epsfig}
\usepackage{xcolor}
\usepackage[normalem]{ulem}

\def\e{{\rm e}}

\def\d{\partial}

\newcommand{\be}{\begin{equation}}
\newcommand{\ee}{\end{equation}}
\newcommand{\bea}{\begin{eqnarray}}
\newcommand{\eea}{\end{eqnarray}}
\newcommand{\bg}{\begin{gather}}
\newcommand{\eg}{\end{gather}}
\newcommand{\bseq}{\begin{subequations}}
\newcommand{\eseq}{\end{subequations}}

\renewcommand{\ln}{\mathop{\rm ln}\nolimits}

\begin{document}
\begin{flushright}
INR-TH-2022-013
\end{flushright}
\vspace{10pt}
\begin{center}
  {\LARGE \bf On the strong coupling problem in cosmologies with
  ``strong gravity in the past''}\\
\vspace{20pt}
Y.~Ageeva$^{a,b,c,}$\footnote[1]{{\bf email:}
    ageeva@inr.ac.ru}, P. Petrov$^{a,}$\footnote[2]{{\bf email:}
    petrov@inr.ac.ru}\\
\vspace{15pt}
  $^a$\textit{
Institute for Nuclear Research of
         the Russian Academy of Sciences,\\  60th October Anniversary
  Prospect, 7a, 117312 Moscow, Russia}\\
\vspace{5pt}
$^b$\textit{Department of Particle Physics and Cosmology,\\
  Physics Faculty, M.V.~Lomonosov
  Moscow State University, \\Leninskie Gory 1-2,  119991 Moscow, Russia
  }\\
\vspace{5pt}
$^c$\textit{
  Institute for Theoretical and Mathematical Physics,\\
  M.V.~Lomonosov Moscow State University,\\ Leninskie Gory 1,
119991 Moscow,
Russia
}
    \end{center}
    \vspace{5pt}
\begin{abstract}
%
 We examine the potential strong coupling problem at early times in a 
 bouncing cosmological
 model with ``strong gravity in the past''
 (Jordan frame), which is conformally related to 
 inflation (Einstein frame). From naive dimensional analysis in the
 Jordan frame
 one would
 conclude that the quantum
 strong coupling energy scale can be lower than the classical
 energy scale.
 However, from the Einstein frame prospective this should not
 be the case. We illustrate this point by calculation in the Jordan frame
 which shows cancellations of the dangerous contributions in the tree level
 amplitude.
\end{abstract}    
    
\section{Introduction}

In scalar-tensor gravities, there is a possibility of bouncing or genesis
 cosmology in
the Jordan frame, with the
effective Planck mass depending on time and tending to zero in the
asymptotic past (``strong gravity in the past'').
Such a scenario has been discussed~\cite{Kobayashi:2016xpl,Ijjas:2016vtq,Ageeva:2020buc, Ageeva:2021yik}, in particular,
in the context of Horndeski theories~\cite{Horndeski:1974wa, Fairlie:1991qe, Nicolis:2008in, Deffayet:2010qz, Kobayashi:2010cm},
where it has been proposed to avoid instabilities otherwise guaranteed by a no-go
theorem~\cite{Kobayashi:2016xpl,Libanov:2016kfc}.
Similar situations have been considered in
other contexts, see, e.g.,
Ref.~\cite{Wetterich:2021cyp} and references therein.

Once the effective Planck mass tends to zero in the asymptotic past,
one may worry that the theory is in the strong coupling regime at early
times, so the classical treatment of the background is not
legitimate. Whether or not this is the case depends on
the relationship between the quantum strong
coupling energy scale and  the classical scale
determined by the Hubble parameter and its time derivatives.
One way to approach this issue is to make use of naive dimensional
analysis of the interacting theory~\cite{Ageeva:2020buc,Ageeva:2018lko,Ageeva:2020gti}. The purpose of this note is
to point out that naive dimensional analysis may sometimes badly fail
in estimating the strong coupling scale.

Our example is the bouncing Universe in the
Jordan frame which is
conformally related to the inflationary Universe in the Einstein
frame~\cite{Nandi:2020sif}.
For an appropriate inflationary scalar potential, the Einstein frame
picture guarantees that there is no strong coupling
problem, i.e., the classical treatment of the background is
fully legitimate. We will see that,
on the other hand, the naive dimensional analysis
in the Jordan frame would show the opposite. This is the problem of
the naive dimensional analysis, however: our direct calculation
of tree level amplitude in the Jordan frame
shows strong cancellations yielding 
consistency
with the Einstein frame inflationary considerations.

We introduce the model in Sec.~\ref{sec:model}, derive the action for scalar
perturbations in the Jordan frame at quadratic and cubic orders in
Sec.~\ref{sec:actions}, 
consider the strong coupling
issue at the level of naive dimensional analysis in Sec.~\ref{sec:naive}
and  finally calculate the tree level amplitude in Sec.~\ref{sec:amplitude}.
We conclude in Sec.~\ref{sec:conclude}.

\section{Bounce conformally related to inflation}
\label{sec:model}

\subsection{Actions}

Following Ref.~\cite{Nandi:2020sif}, we consider
a class of bouncing models (Jordan frame)
that are conformally related 
to cosmological inflation. 
The action in the Jordan (bounce)
frame is given by 
\begin{equation}
\label{horndeski_act_bounce}
    \mathcal{S}_{b} = \int d^4 x \sqrt{-g}\left[ P(\phi, X)
    + \frac{M_{P}^2 f^2(\phi)}{2} R\right]\; , 
\end{equation}
with
\begin{equation*}
P(\phi, X) = \omega(\phi)X-V(\phi),
\end{equation*}
where
  $M_P = (8\pi G)^{-1/2}$ is  reduced Planck mass, 
$R$ is  Ricci scalar  and
\begin{equation*}
    X = -\frac{1}{2}g^{\mu v} \partial_{\mu} \phi \partial_{\nu} \phi,
\end{equation*}
\begin{equation*}
\omega(\phi)=f^{2}-6 M_{P}^{2} \left(\frac{d f}{d\phi}
  \right)^2  \; , \;\;
\quad V(\phi)=f^{4}(\phi) V_{I}(\phi)\; .
\end{equation*}
Here $f(\phi)$ is a yet undetermined function, and $V_I(\phi)$
is the scalar potential in the Einstein frame.
We do not use special notation for quantities in the Jordan frame;
notations here agree with Ref.~\cite{DeFelice:2011zh}, modulo
definition $F(\phi) = f^2 (\phi)$.

By conformal transformation
\be
\nonumber
g_{\mu \nu} = f^{-2}(\phi) g_{I\, \mu \nu}
\ee
the theory \eqref{horndeski_act_bounce} is related to the 
following inflationary model in the
Einstein (inflation) frame:
\begin{equation*}
S_I=\frac{1}{2} \int d^4 x \sqrt{-g_I}\left[M_{P}^2R_I
-g_I^{\mu v} \partial_{\mu} \phi \partial_{\nu} \phi-2 V_I(\phi)\right],
\end{equation*}
where subscript ``I'' refers to quantities in the 
Einstein frame.

\subsection{Einstein frame: inflation}

We consider inflation potential that flattens out
at large fields,
 \begin{equation}
   V_I(\phi) \to V_\infty,  \quad \text{as} \quad \phi\to \infty \; ; \;\;\;
   \;\;\;\; V_{\infty} \ll M_P^4 \; ,
   \label{jun15-22-1}
 \end{equation}
 so that the energy density is always sub-Planckian.
 Viewed from the Einstein frame, 
 the classical description of inflating background and semiclassical
 treatment of cosmological perturbations are perfectly legitimate.
 Inflation occurs in the slow roll regime at early times,
 $\epsilon \ll 1$, $\eta \ll 1$,
 where we use the standard notations
\be
   \epsilon = \frac{(V_{I}')^2 M_P^2}{2 V^2}
\; , \;\;\;\;
   \eta = \frac{V_{I}^{\prime \prime} M_P^2}{V}\; .
  \label{jun17-22-10}
  \ee
 The slow roll equations are

 \be
 \label{jun13-22-100}
    \frac{d\phi(\tau)}{d\tau} =  
    -\frac{M_{P}V'_{I}}{\sqrt{3V_I}} \; , \;\;\;\;\;
    H_I = \sqrt{\frac{V_I}{3}}\frac{1}{M_{P}}\; ,
\ee
    where $\tau$ is cosmic time in the Einstein frame.

\subsection{Jordan frame: bounce}

We follow Ref.~\cite{Nandi:2020sif}
 and choose the function defining
the conformal
transformation as follows:
\begin{equation*}
f(\phi)=f_0 \text{exp}\left[-\frac{(\alpha+1)}{M_{P}^2}\int 
d\phi \frac{V_I}{V_{I}'}\right],\quad \alpha>0,
\end{equation*}
where the value of $f_0$ is irrelevant for our purposes.
Then the Jordan frame metric is
\be
\nonumber
ds^2 = f^{-2} (\phi(\tau)) d\tau^2 -   f^{-2} (\phi(\tau)) a^2_I (\tau)
d{\bf x}^2
\ee
and the Hubble parameter in the Jordan frame is given by
\be
H =  f \frac{d}{d\tau} \ln (a_I f^{-1}) =
-f \cdot \frac{\alpha}{M_P} \sqrt{\frac{V_I}{3}} \; ,
\label{jun14-22-1}
\ee
where we make use of the slow roll equations \eqref{jun13-22-100}.
The Jordan frame universe contracts (and, at the end of the
Einstein frame inflation, experiences the bounce).

It is worth emphasizing that $f \to 0$ as $t\to -\infty$.
So, the Jordan frame Hubble parameter vanishes in the asymptotic past.
The Jordan frame effective Planck mass $M_P^{(eff)}=f M_P$ also tends to zero
as  $t\to -\infty$; this situation is dubbed ``strong gravity
in the past''.

\section{Strong coupling and absence thereof}
\label{sec:strong_coupling}

From now on we work in the Jordan frame.

We concentrate on the scalar sector of perturbations about the contracting
solution \eqref{jun14-22-1}. We use the unitary gauge
\be
\nonumber
\delta \phi = 0 \; ,
\ee
then
the scalar perturbation is parameterized with the field $\zeta$,
entering the spatial metric, so that the full metric in the Jordan frame cosmic
time is
\cite{DeFelice:2011zh}
\begin{equation*}
    ds^2 = -[(1+\alpha)^2 - a^{-2} \text{e}^{-2\zeta}(\partial\psi)^2]dt^2 
    + 2\partial_i\psi dt dx^i + a^2 \text{e}^{2\zeta} d{\bf x}^2 \; ,
\end{equation*}
where $\alpha $ and $\psi$ are perturbations of the lapse and shift.
Upon solving the constraints, one arrives at the unconstrained action
written in terms of $\zeta$.
We consider its quadratic and cubic parts. To this end, we  adapt the
results of Ref.~\cite{DeFelice:2011zh}.

\subsection{Quadratic and cubic  actions}
\label{sec:actions}
The  quadratic action for scalar 
perturbation is 
\begin{equation*}
\mathcal{S}^{(2)}_{\zeta\zeta}=\int d t d^{3} x a^{3} \mathcal{G}_S\left[\dot{\zeta}^{2}-\frac{1}{a^{2}}\zeta_{,i}\zeta_{,i}\right],
\end{equation*}
where, using  
formulas
given in Ref.~\cite{DeFelice:2011zh,Gao:2012ib}, we obtain
\be
\nonumber
\mathcal{G}_S  = \frac{1}{2} \frac{\dot{\phi}^2}{H_I^2} = \frac{f^2}{2 H_I^2}
\left(\frac{d\phi}{d\tau} \right)^2 \; .
\ee
This is an exact expression, which is actually a straightforward
Jordan frame reformulation 
of the standard Einstein frame result. In the slow roll case
\eqref{jun13-22-100} one has
\begin{align*} 
 \mathcal{G}_S  = f^2\cdot\frac{M_{P}^4 (V'_I)^2}{2V_I^2} \; .
 \end{align*}
Note that the perturbations propagate luminally, which is again a
Jordan frame counterpart of the standard Einstein frame property.

The terms in the cubic action for scalars, which
do not vanish in the model \eqref{horndeski_act_bounce} either identically
or due to background equations and field
redefinition~\cite{DeFelice:2011zh,Gao:2012ib,Maldacena:2002vr}, are
(we use notations of Ref.~\cite{DeFelice:2011zh})
\begin{align}
\label{cubic}
\mathcal{S}^{(3)}= \int dt d^{3}
x~a^3 \left\{ \mathcal{C}_{1}  \zeta \dot{\zeta}^{2}
+\frac{1}{a^2} \mathcal{C}_{2} \zeta(\partial \zeta)^{2}
+\mathcal{C}_{4} 
\dot{\zeta}\left(\partial_{i} \zeta\right)\left(\partial_{i} \mathcal{X}\right)
+ \mathcal{C}_{5}  \partial^{2} 
\zeta(\partial \mathcal{X})^{2}
\right\}\; ,
\end{align}
where $\d^2 = \d_i \d_i$ and
$\partial^2\mathcal{X} =  \dot{\zeta}$.
The coefficients are straightforwardly calculated. To the leading order in
the slow roll parameters we have
\begin{subequations}
  \label{jun15-22-2}
\begin{align}
\mathcal{C}_{1}&= f^2\cdot \frac{M_{P}^6(V'_{I})^2}{4V_{I}^4}\big(4 V_I V''_I
- 3 (V'_{I})^2\big), \\
\mathcal{C}_{2}&=  f^2\cdot \frac{M_{P}^6(V'_{I})^2}{4V_{I}^4}\big(5 (V'_{I})^2
- 4 V_I V''_I \big), \\
\mathcal{C}_{4}&= f^2 \frac{M_{P}^6(V'_{I})^4}{16V_{I}^6}\big(M_{P}^2 (V'_I)^2
-8 V_{I}^2  \big), \\
\mathcal{C}_{5}&= f^2  \frac{M_{P}^8 (V'_{I})^6}{32 V_I^6} \; .
\end{align}
\end{subequations}

\subsection{Naive dimensional analysis}
\label{sec:naive}
We now proceed with the naive dimensional analysis of the
strong coupling problem.
The classical energy scale is of order of the Hubble parameter
\eqref{jun14-22-1},
\begin{equation}
\label{class_1}
    |E^{(class)}| = |H| \sim  \frac{f\sqrt{V_I}}{M_P} \; .
\end{equation}
To obtain an estimate of the strong coupling scale through naive
dimensional analysis, we set, at a given moment of time, $a=1$
and introduce 
canonically normalized field 
\begin{equation*}
    \zeta_c = \sqrt{2\mathcal{G}_S} \zeta \; . 
\end{equation*}
In terms of the canonically normalized field,
the cubic action still has the form \eqref{cubic}
with the replacement
\be
\nonumber
\tilde{\mathcal{C}}_{i} =  (2\mathcal{G}_S)^{-3/2} {\mathcal{C}}_{i} \; ,
\ee
so that
\begin{align*}
    \tilde{\mathcal{C}}_1 &
    = \frac{1}{f}\cdot\frac{(-3 (V'_I)^2+4 V_I V''_I)}{4 V_I V'_I} ,  \\
    \tilde{\mathcal{C}}_2 &
    = \frac{1}{f}\cdot\frac{(5 (V'_I)^2-4 V_I V''_I)}{4 V_I V'_I}.
\end{align*}
while
\be
\tilde{\mathcal{C}}_{4} \sim  \frac{1}{f} \cdot \frac{V_I'}{V_I} \; , \;\;\;\;\;
\tilde{\mathcal{C}}_{5} \sim  \frac{1}{f} \cdot M_P^2
\left(\frac{V'_I}{V_I} \right)^3 \; .
\label{jun17-22-1}
\ee
All operators in the resulting cubic Lagrangian are dimension-5,
so one immediately finds naive estimates for the associated
strong coupling scales,
\be
\nonumber
E_{ i}^{(naive)} \sim |\tilde{\mathcal{C}}_{i}|^{-1} \; .
\ee
Naively, the most relevant of these scales are the lowest ones,
which are associated with the largest ${\mathcal{C}}_{i}$.

For asymptotically flat inflaton potential \eqref{jun15-22-1}, one typically has
$\eta \gg \epsilon$, so the largest couplings in \eqref{jun15-22-2} are
$\mathcal{C}_{1}$ and $\mathcal{C}_{2}$. The two naive strong coupling
scales are of the same order:
\begin{equation}
\label{strong_1}
E^{(naive)} \sim f \frac{V_I'}{V''_I}.
\end{equation}
Depending on the shape of the inflaton potential, classical energy  scale \eqref{class_1} may
exceed strong coupling energy scale
\eqref{strong_1}. As an example, for the inflaton potential
\be
\nonumber
V_I =  V_\infty \left(1 - \e^{\phi^2/\mu^2}\right)
\ee
one has
\be
\nonumber
\frac{E^{(naive)}}{ E^{(class)}} \sim \frac{\mu^2}{\phi H_I} 
\ee
which is less than 1 at large $\phi$.

We conclude that naive dimensional analysis in the Jordan frame
suggests that
there is a  quantum strong coupling
 energy scale which, for appropriate inflaton potential,
is below the classical scale. If not for the Einstein frame
considerations, one would be tempted to dismiss such a model.

To end up this Section, we notice that the cubic couplings
$\tilde{\mathcal{C}}_{4}$ and  $\tilde{\mathcal{C}}_{5}$ are
not enhanced, see \eqref{jun17-22-1}.
So,  at large $\phi$, their associated strong coupling scales
are much higher than the
classical energy scale \eqref{class_1}.
In other words, the third and fourth  terms in
the integrand in \eqref{cubic} per se do not imply
strong coupling, even naively.
Thus, we do not have to consider the terms with couplings
$\mathcal{C}_{4}$ and  $\mathcal{C}_{5}$ in our
analysis of the amplitudes.

\subsection{Scattering amplitude}
\label{sec:amplitude}
Making use of the first and second terms in the cubic action
\eqref{cubic}, with  $\mathcal{C}_{1,2}$  replaced by
$\tilde{\mathcal{C}}_{1,2}$ and $\zeta$ by canonically normalized
$\zeta_c$, it is straightforward to calculate $2 \to 2$
scattering amplitude. Before giving the result, we note that
if we set, for the sake of argument, $\tilde{\mathcal{C}}_{2} = 0$,
then the matrix element would be given by
\begin{equation*}
  M_{\tilde{\mathcal{C}}_1\, ; \, \tilde{\mathcal{C}}_2 = 0}
  = -\frac{E^2}{f^2}\cdot \frac{(9x^2-5)\big(3(V'_I)^2
    - 4V_IV''_I\big)^2}{64(x^2-1)V_I^2(V'_I)^2} \; ,
\end{equation*}
where $x=\cos \theta$ and $\theta$ is scattering angle.
Were this the correct matrix element,
our naive expectation would be confirmed: the partial wave amplitudes
\be
\nonumber
a^{(l)} = \frac{1}{32\pi}\int~dx~P_l(x)\,
M_{\tilde{\mathcal{C}}_1\, ; \, \tilde{\mathcal{C}}_2 = 0}
\ee
where $P_l$ is the Legendre polynomials,
would hit the unitarity bound $|a^{(l)}| = 1/2$ at $E \sim E^{(naive)}$. 
The same situation would occur if we set  $\tilde{\mathcal{C}}_{1} = 0$.

However, there are strong cancellations. Indeed, the matrix elements
in $s$-, $t$- and $u$-channels are, respectively
\begin{subequations}
\begin{align*}
 M_s &= -\frac{E^2}{4} (3\tilde{\mathcal{C}}_1
 +\tilde{\mathcal{C}}_{2})^2 \; ,
 \\
   M_t &= \frac{E^2}{2(1-x)}
    \Big[\tilde{\mathcal{C}}_1 + \tilde{\mathcal{C}}_2 (2-x) 
      \Big]^2 \; ,
    \\
    M_u &= \frac{E^2}{2(1+x)}
    \Big[\tilde{\mathcal{C}}_1 + \tilde{\mathcal{C}}_2 (2+x) \Big]^2 \; .
\end{align*}
\end{subequations}
The resulting matrix element is
\begin{equation*}
    M = M_s + M_t + M_u = \frac{E^2}{f^2}\cdot \frac{(41x^2-45)(V'_I)^2 
    - 40(x^2-1)V_IV''_I}{16(x^2-1)V_I^2} \; .
\end{equation*}
We see that the strong coupling scale is actually given by\footnote{We
  still consider the
case $V_IV_I'' \gg (V_I')^2$; we cannot trust the term with $(V_I')^2$
in the numerator anyway, since we neglected terms with ${\mathcal{C}}_{4,5}$
in the cubic action \eqref{cubic}.}
\be
\nonumber
E^{(strong)} \sim f \cdot \left(\frac{V_I}{V_I''} \right)^{1/2} \sim
f\cdot \frac{M_P}{\eta^{1/2}} \; ,
\ee
where $\eta$ is the slow roll parameter \eqref{jun17-22-10}. As anticipated,
this scale
is much higher than the classical energy scale \eqref{class_1} for
$V_I \ll M_P^4$. Our calculation of the amplitude confirms
the absence of the strong coupling problem.

\section{Conclusion}
\label{sec:conclude}

Of course, the model we have considered in this paper
is nearly trivial.
Still, it illustrates the main point: naive dimensional analysis
may grossly underestimate the quantum strong coupling energy scale.
There may be less trivial  situations where
this property holds,
 e.g., due to kinematical or dynamical symmetries.
It would be interesting to have more examples and see whether
the mismatch between the dimensional analysis and actual
strong coupling scale can always be understood via
field redefinitions.

\section*{Acknowledgments}
The authors are grateful to Valery Rubakov for useful comments and fruitful discussions as well as for  careful reading of the early versions of this manuscript.  This work has been supported by Russian Science Foundation Grant No. 19-12-00393.

\end{document}